**TITLE:** Pathologists light level preferences using the microscope - a study to guide digital pathology display use

**AUTHORS:** Charlotte Jennings[1,2], Darren Treanor[1,2,3], David Brettle[1]

**AFFILIATIONS:**
1) National Pathology Imaging Co-operative, Leeds Teaching Hospitals NHS Trust, Leeds, UK
2) Section of Pathology and Data Analytics, Leeds Institute of Medical Research, University of Leeds, Leeds, UK
3) Centre for Diagnostics, Division of Neurobiology, Department of Clinical and Experimental Medicine, Department of Clinical Pathology, Linköping University, Linköping, Sweden

**CORRESPONDING AUTHOR:**
Dr Charlotte Jennings
Email: charlotte.jennings1@nhs.net

**ABSTRACT:**

**Background**

Currently there is a paucity of guidelines relating to displays used for digital pathology making procurement decisions, and optimal display configuration, challenging.
Experience suggests pathologists have personal preferences for brightness when using a conventional microscope which we hypothesised could be used as a predictor for display setup.

**Methods**

We conducted an online survey across 6 NHS hospitals, totalling 108 practising pathologists, to capture brightness adjustment habits on both microscopes and screens.

A convenience subsample of respondents were then invited to take part in a practical task to determine microscope brightness and display luminance preferences, in the normal working environment.
A novel adaptation for a lightmeter was developed to directly measure the light output from the microscope eyepiece.

**Results**

The survey (response rate 59% n=64) indicates 81% of respondents adjust the brightness on their microscope. In comparison, only 11% report adjusting their digital display. Display adjustments were more likely to be for visual comfort and ambient light compensation rather than for tissue factors, common for microscope adjustments. Part of this discrepancy relates to lack of knowledge of how to adjust displays and lack of guidance on whether this is safe; But 66% felt that the ability to adjust the light on the screen was important.

Twenty consultants took part in the practical brightness assessment. Light preferences on the microscope showed no correlation with screen preferences, except where a pathologist has a markedly brighter microscope light preference.  All of the preferences in this cohort were for a display luminance of less than 500cd/m$^2$, with 90% preferring 350cd/m$^2$ or less. There was no correlation between these preferences and the ambient lighting in the room.

**Conclusions**

We conclude that microscope preferences can only be used to predict screen luminance requirements where the microscope is being used at very high brightness levels. A display capable of a brightness of 500cd/m$^2$ should be suitable for almost all pathologists with 300cd/m$^2$ suitable for the majority. Although display luminance is not frequently changed by users, the ability to do so was felt to be important by the majority of respondents.
Further work needs to be undertaken to establish the relationship between diagnostic performance, luminance preferences and ambient lighting levels.

**INTRODUCTION:**

There is a rapid increase in the clinical use of digital pathology internationally. The promise of improved workflows, better connectivity between pathologists and providing services to remote locations are the driving force for this change and digital pathology is frequently cited as part of the solution to address an international shortage in the pathology workforce [1]. The FDA granted licensing for the first clinical digital pathology system in 2017, which provided regulatory support and thus facilitated policy makers through to pathology departments in the push to "go digital" [2].

However, for health care services going digital requires a significant initial financial outlay, which was reported to be a key barrier to adoption in a 2018 survey of UK pathology departments [3]. Part of this expense includes the procurement of the relevant hardware required for digital pathology workflows and includes the displays on which pathologists will report whole slide images [4]. A vast array of displays are available ranging across medical grade, consumer-off-the-shelf and professional models, with widely varying associated costs [5,6]. Navigating the many described specifications of these displays is challenging but key parameters have been proposed to include luminance, contrast, colour accuracy, resolution and "just noticeable difference" [6,7]. Currently, there is no consistent guidance about the specifications required for these displays as illustrated by Chong et al where the minimum requirements for a range of national guidelines cover a screen size range of 17 to 27" and a luminance maximum of 100 to 300 $cd/m^2$ [8]. Williams et al recommended a minimum specification of 24-27" and 250-350 $cd/m^2$ in guidance for remote reporting on pragmatic grounds [9].

But learning from experience in radiology, guidelines will be necessary to ensure that the technical performance of these displays is sufficient as well as to address ergonomic aspects of screen working [8,10–12].

This uncertainty is reflected in the approach taken by the US Food and Drugs Administration (FDA) to grant clinical approval to a whole system, with a pre-defined display [13]. However, during the COVID pandemic, the drive to enable remote working (on different displays) was met with a statement from the FDA that pathologists should "use their clinical judgement to determine whether the quality of the images…are sufficient for interpretation" and indicates a move towards pathologists having a responsibility to ensure their displays are fit for purpose, as they do for their microscopes [6].

In 2018, our pathology department (Leeds Teaching Hospitals NHS Trust, UK) implemented 100% digital scanning of all slides. As part of this process, consultant pathologists were supplied with large, medical-grade, high resolution and high luminance displays which were chosen based on best available evidence (Jusha, Nanjing Jusha & Commercial Trading Ltd, China: Model C620L with 6 megapixel 30-inch display and luminance up to 800$cd/m^2$), as summarised in the Leeds Guide to Digital Pathology Vol 1 [14,15]. A minority of users struggled with the perceived "brightness" of these displays and were developing symptoms of visual strain. In attempting to tailor the display to these users as per Government Health and Safety Display Screen Equipment guidance a conflict arose in altering the medical-grade monitor from its approved settings [16]. Similar conflicts between screen parameters and visual strain were experienced in radiology and have largely been navigated through ambient light control and ergonomic working practices and have again been incorporated into guidance documents [10].

In contrast to radiology, current pathology practice does not involve the regulation of ambient lighting and in the UK consultants generally work in individual use offices where further variation in ambient light is introduced by individual preference [10,17]. This makes sense given that the closed nature of a microscope is far less likely to be impacted by ambient lighting. Furthermore, microscopes allow the user to make easy adjustments to the lighting by a continuous dial which allows for adjustment for individual sensitivities to light as well as to navigate tissue factors such as thick sections. Anecdotally, there is significant variation between pathologists in their use of microscope brightness levels, perhaps especially noticed by trainee pathologists who are more likely to use double-header microscopes to review images and to rotate between a number of microscopes, left at the setting of the previous user, during their training. Deploying digital pathology systems in such variable ambient conditions and for pathologists who are used to sensitive and easy control of light while they work presents an additional challenge for departments in setting up displays.

We conducted a survey of 6 pathology departments in the UK to understand perspectives on, and variation of, light use at both the screen and microscope. We then designed a practical experiment to capture the working light preferences of pathologists at both the microscope and screen to test our hypothesis that microscope light preferences would correlate with screen preferences and could be used as a predictor for display set up.

**METHODS:**

This study comprised two phases: An online survey circulated via email and a practical assessment of light preference.

<u>Survey of light use habits and preferences</u>

An online survey was designed to capture the light use habits and preferences of pathologists across the West Yorkshire Association of Acute Trusts Region in the UK. This region includes 108 pathology consultants and trainees, working across 6 different NHS trusts. At the time of writing the pathologists in these trusts were in the middle of a region wide deployment of digital pathology and have differing experiences of digital pathology. This ranges from departments with limited or no experience of digital pathology to a department with a fully enabled digital workflow.

The survey was conducted through Microsoft Forms and shared via email to a lead pathologist at each trust who circulated the survey to their teams. The survey was composed of 12 questions and captured limited demographic data, such as age and role, as well as information about current light use habits on microscope and screen. General opinions about the importance of light adjustability were also invited in a free text question.
Three reminder emails were sent over a period of 8 weeks before the online form was closed to further responses. A copy of this questionnaire is available for review in *Supplementary material S1*.

<u>Assessment of light preferences</u>

A subset of the main cohort was invited to take part in a practical measurement of their light preferences at the microscope and digital displays. This cohort comprised 40 consultants at a large tertiary teaching hospital (Leeds Teaching Hospitals NHS Trust), which underwent a full workflow digitisation in 2018. All consultants in this centre have high specification medical grade displays (Jusha C620L) selected for digital reporting as part of the digital deployment programme which uses a Sectra PACS slide viewing software. The consultant microscopes are of varying make, model and age. Within the department there is mixed digital pathology experience and competence, ranging from minimal to full time use for primary reporting and other functions.

The experiment entailed a light preference adjustment task performed on slides and digital images in the participants usual workspace on their usual reporting equipment. Preference was defined for participants as "*a point where you feel visually comfortable and also feel able to assess features of the slide or image at a level that is needed for reporting*". One to one appointments were made with author CJ to conduct the experiment over the 2-month period of the experiment.

An LXCan spot luminance meter (IBA international, Louvain-La-Neuve, Belgium. Within manufacturer calibration at the time of use) was used to measure ambient lighting/illuminance (lux) and screen luminance ($cd/m2$). Light output directly from the microscope eyepiece was also measured using the illuminance detector via the development of a novel adaptation *(described in Supplementary material S2)*. Slides of two different tissue types (breast and liver) were prepared in the departments own laboratory following usual haematoxylin and eosin staining protocols and converted to digital images on a Leica GT450 (Leica Biosystems, Nußloch Germany), the scanner normally used in the department. Corresponding areas on the slides and images were marked to direct the pathologists to a specific region of the tissue and control at what magnification it would be assessed (*Figure 1*).

A scripted protocol was used from the consent process through to completion of the experiment to standardise data collection.

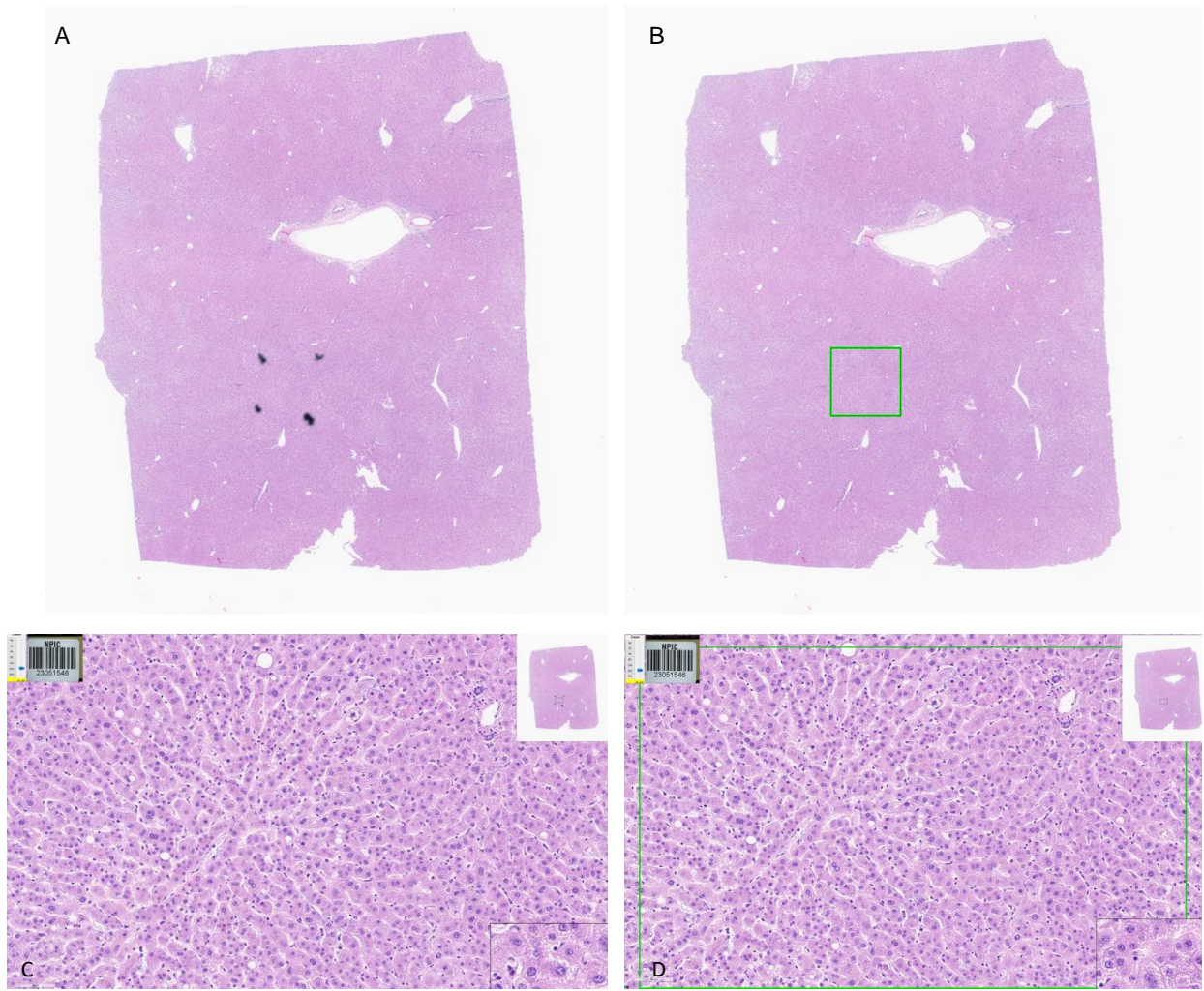

*Figure 1. Marked up (pen) area of slide (A) and marked up image (B) of the section of liver parenchyma used in demonstration of the practical task. Pathologists were advised to centre their view at the middle of the markings and increase the objective lens until the marking was just out of view (microscope) or the digital magnification until the green box was at the edge of the screen (screen). This set the viewing magnification on both modalities at 20x (C and D).*

**Set up**

Participants were first asked to make any adjustments to their person, space or equipment that they would normally make prior to reporting. Ambient lighting and equipment position was recorded at this point.

**Screen preference**

The pre-experiment screen luminance was recorded on a demonstration pathology image (whole slide image of liver parenchyma scanned at 40x magnification (0.26 microns per pixel) on a Leica GT450 scanner) which participants were asked to centre on the pre-defined region of the image.

The backlight was then reduced to its minimum by the experiment controller. As the light was then gradually increased, participants were asked to state when the light level had reached their preference for viewing the image. This process was then repeated three times with a test slide (a whole slide image of breast tissue containing invasive carcinoma).

**Microscope preference**

The same approach was taken on the microscope, taking a pre-experiment measurement, working through a demonstration slide (liver parenchyma) and then repeating three times on the test slide (breast tissue). On the microscope, the experiment controller reduced the microscope light dial to its minimum before the experiment began but the participants were invited to control the dial when choosing their preference. Experiment set up pictured in *Figure 2.*

At the end of the experiment, participants were invited to comment on any thoughts about light use on their equipment generally or thoughts about the experiment. Comments from during the experiment were also noted.

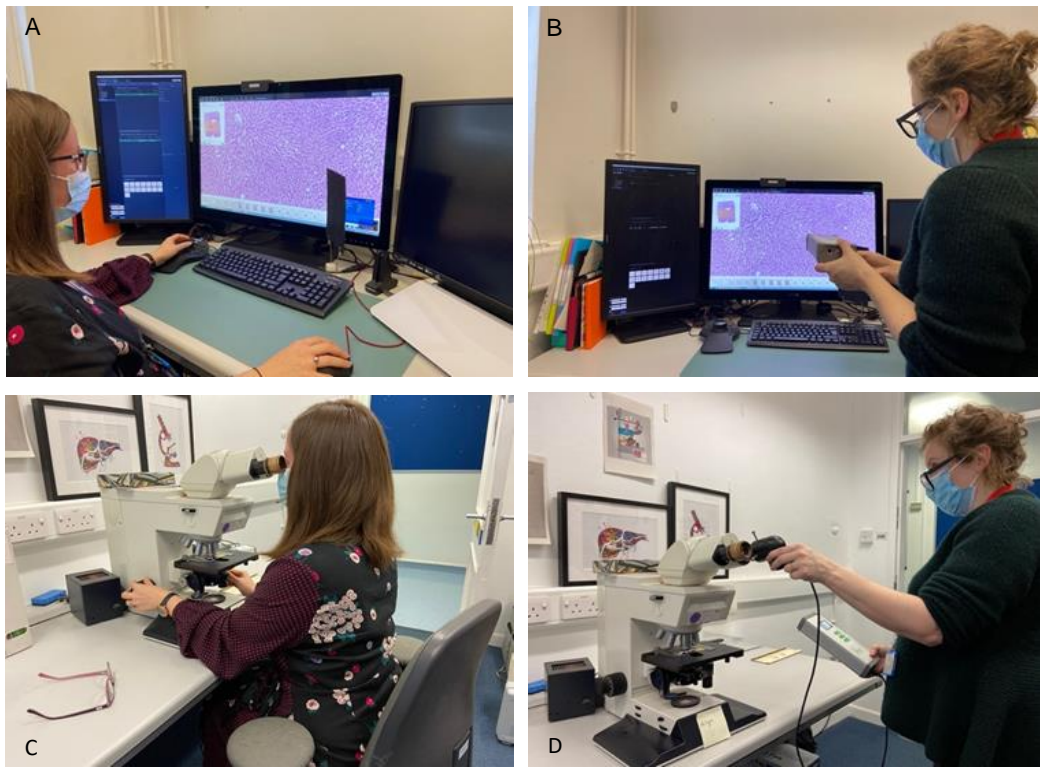

*Figure 2. Images of the experiment set up for pathologist assessment (A) and measurement (B) on the digital display and assessment (C) and measurement (D) at the microscope.*

**RESULTS:**

Online Survey

The online survey was completed by 64 pathologists, a response rate of 59%. The respondents included 52 consultants (65%) and 12 trainees (43%) and covered the full working age range of the cohort (*Figure 3*). A variety of different work patterns were represented, with participants spending between 1 and 63.5 hours reporting per week.

As expected, there was quite variable usage of digital pathology. Time spent viewing digital images ranged from 0.5 to 25 hours and viewing slides on a microscope ranged from 1 to 50 hours with several pathologists using a combination of modalities (*Figure 3*). Twenty-three percent describe no use of digital pathology at all. While for 15% primary reporting was their predominant usage, the most prevalent use of digital pathology was for teaching and training (45%), with smaller numbers using for MDT (12.5%) and secondary opinions (3.1%).

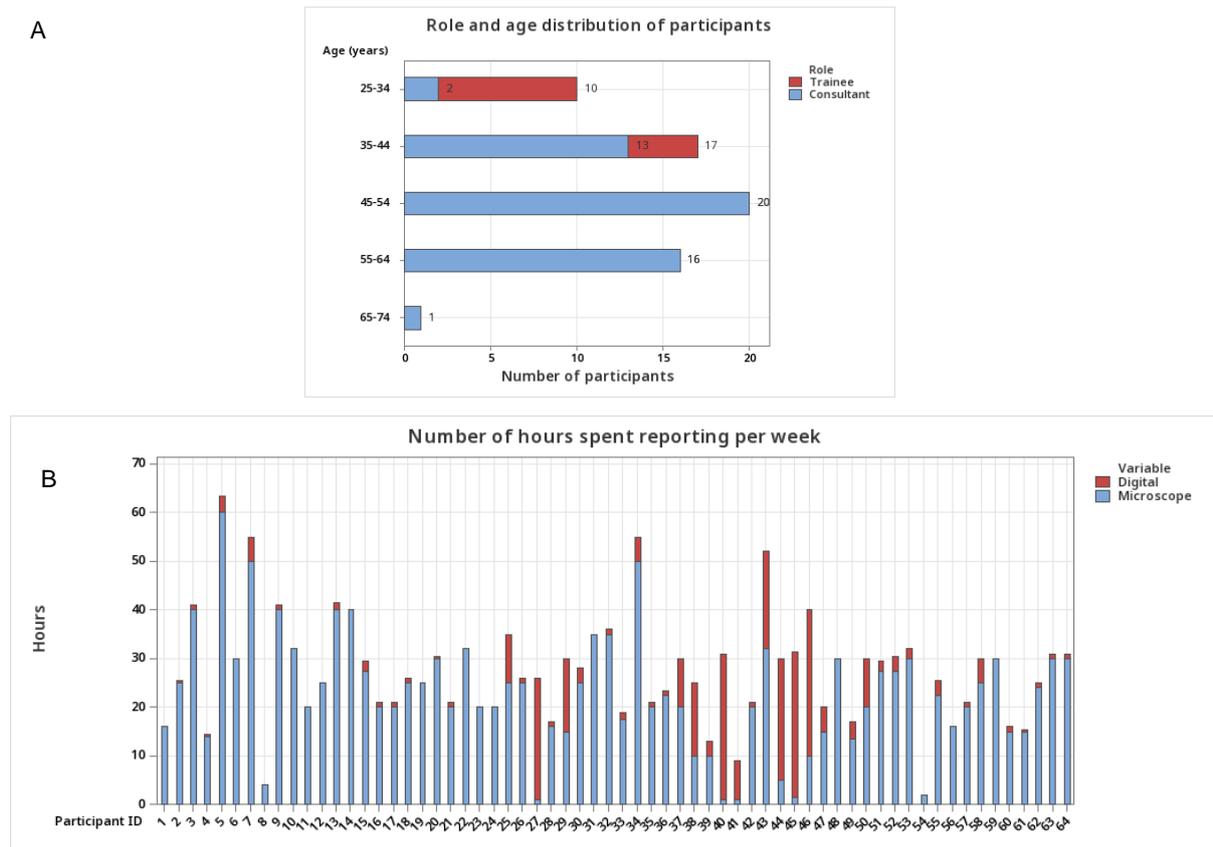

*Figure 3.* Participating pathologists split by their (A) role and age, and (B) number of hours and modality spent on reporting per week.

**Light use habits**

Adjustment of light at the microscope was common, with 81% of our respondents adjusting the light on some occasions. The frequency of light adjustment covered the full spectrum of the scale that was presented [1= never, 5 = half of reporting sessions, 10 = every case]. Most of the participants (73%) estimate adjusting microscope light settings half of the times they report or less. However, one respondent reported making light adjustments for every case they view (*Figure 4*).

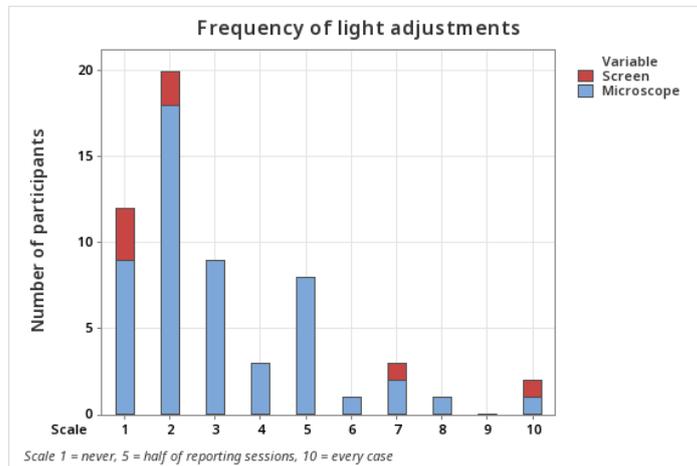

*Figure 4.* Frequency of light adjustments to reporting equipment by survey participants. Screen adjustments made by 7 participants and 52 microscope users.

In comparison, only 11% reported ever adjusting the backlight settings on their digital display. Of those that had adjusted their displays (7 people), the frequency spectrum of adjustments was just as broad, and 1 user also indicated they adjusted the display for every case (possible respondent error).

Participants were able to select multiple options to describe the reasons for light adjustment on their equipment. The options given included adjustments for slide related features (such as for specific diagnostic features, specific stains or tissue types and slide quality issues), viewing at different objectives, changes in environment lighting, for visual comfort or adjustment from another user. All users who adjust their lighting do so for multi factorial reasons. For those who adjust their digital display, visual comfort and changes in environment lighting are the dominant factors. In contrast, at the microscope, slide factors predominate (*Figure 5*).

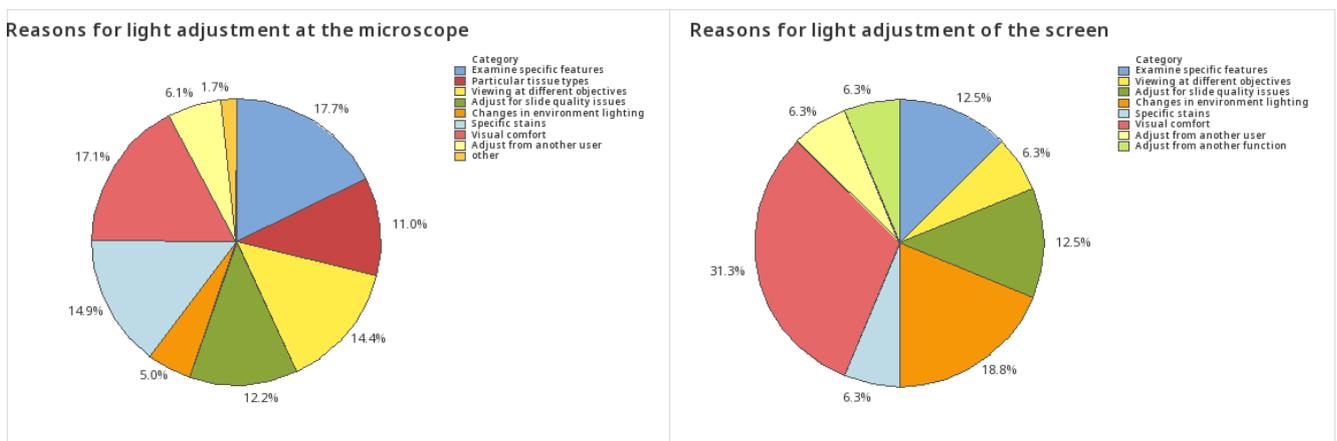

*Figure 5.* Reasons for pathologists' adjustment to the light at (A) the microscope, and (B) their digital display when reporting. Participants were able to select multiple options to best reflect their light use habits.

These differences were further elaborated on in free-text comments, with selected examples provided in Table 1.

| Table 1. Comments relating to light adjustment | |
|---|---|
| **Microscope adjustment** | **Screen adjustment** |
| *"On a light microscope I find it most important to be able to adjust the light when looking into groups of cells on a cytology cytospin preparation, but this wouldn't apply to a digital display."* | *"I think it could be helpful to be able to adjust the light levels to compensate for changes in ambient lighting. This could be more of an issue for digital because you are taking in light from the whole room whilst looking at a screen, whereas when looking down the microscope you move your head so your eyes are very close to the eyepieces and mainly taking in light from the microscope so adjusting the light levels on a microscope might be less important."* |
| *"It is important to be able to adjust the microscope light in certain circumstances (e.g. when polarising), or to adjust for your own viewing if using a shared microscope e.g. multiheader."* | |
| *"Main one is to account for variation in slide thickness with microscope which is a major issue in our lab where technical consistency is unusual"* | *"The room lighting level can easily change and that could affect the screen."* |
| *"The adjustability of the light microscope light level is greater than digital and the ability to adjust the light up for looking at thick sections/dense groups on cytology and for special techniques such as polarising microscopy is essential (particularly as this can't be done digitally)"* | *"I've not really adjusted the backlight on my digital display since starting to use digital having found a light level that doesn't cause eye strain and headaches but is sufficient for reporting. I would see this as the primary reason to be able to adjust the display light as having a brighter display doesn't particularly change the ability to report e.g. thick or overstained sections."* |

**Opinions on light adjustment and displays**

The majority of respondents (59%) found their current displays "neither too bright or too dim" for viewing digital images. Outside the majority a relatively even spectrum existed of those who found the monitor too bright or too dim to some degree. This survey cohort included those who had high specification medical grade monitors as well as those with standard displays (***Figure 6***).

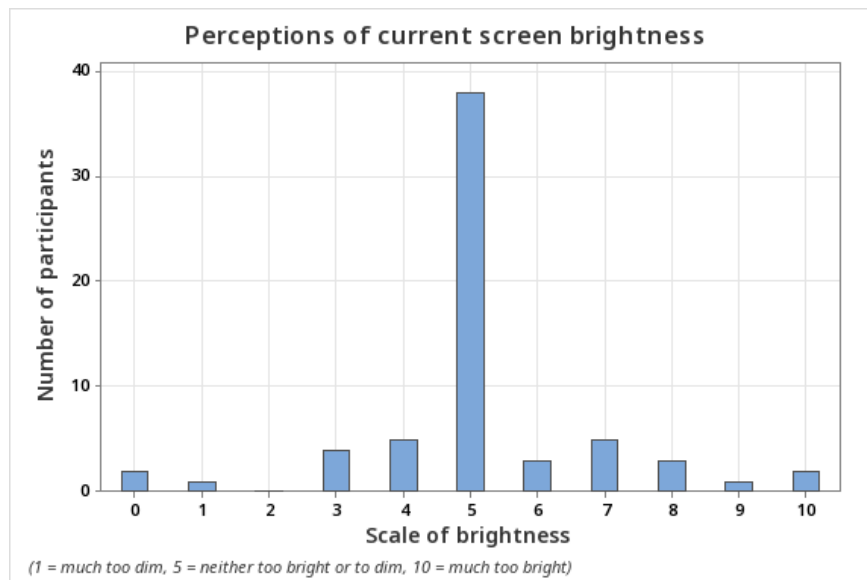

***Figure 6.*** *All survey participants rated their current screen for brightness according to a scale of "much too dim" to "much too bright"*

It was possible to separate the cohort of respondents with high specification medical grade monitors. The majority of this cohort (54%) also found their monitor "neither too bright or too dim" but those outside of this were more likely to find their monitor too bright to some degree.

All of the respondents thought it was important or very important to be able to adjust the light on the microscope. Despite minimal reported adjustment of screens in this cohort, 77% felt it was important or very important to be able to do so, with the remaining 23% saying they were uncertain.

A number of general comments were given to support the importance of adjusting light which covered both reporting factors and user factors, examples in Table 2.

| Table 2 Comments on the importance of light adjustment | |
|---|---|
| **Reporting factors** | **User factors** |
| *"Light is required to allow proper evaluation of contrast."* | *"To be able to suit individual requirements/preferences for brightness."* |
| *"Adjusting the light on a microscope or digital display is important to be able to identify specific features and to examine sections using special stains"* | *"Because no one day or case is like the others."* |
| | *"Our eyes are all different."* |
| *"To get the maximum information from each slide."* | *"Reduce tension headaches, eye strain"* |

A few respondents with no digital pathology experience felt less able to comment on the importance of light adjustment on this modality, for example *"I don't report diagnostic digital slides so I am unsure whether I would use the digital slides in the same way."* However, others felt light adjustment was a core part of reporting *"I don't employ digital pathology, yet......Given how important lighting is in light microscopy, I feel the same thing can be said about digital pathology."*

Free text comments also suggest that the infrequency of screen adjustment seen in this cohort may represent several different issues. Some respondents do not use digital pathology so have no practical need to. Some digital pathology users "*have never thought to do so*", assuming "*once the display is calibrated it should be suitable for displaying all digital images*". Whilst other users "*have never actually figured out how to turn the brightness down! By the end of the day, my eyes tend to get quite fatigued.*" or have concerns about how the changing the settings would affect diagnostic capability "*I am aware that there is some research into potentially missing things on dimmer settings so would be cautious to change without approval/advice*".

Light preference task

An opportunistic sample of 20 consultants took part in the practical light adjustment task, representing 50% of the target cohort.

**The reporting environment**

There was marked variation in the ambient lighting between different participant's offices (Range 39 – 1308 lux), however the majority were operating in normal "office" conditions of around 300 lux (Graph 9). Options for lighting adjustment within different offices was variable, including access to and use of blinds, additional lighting (eg. Uplight lamps and desk lamps) and number and position of overhead bulbs. Display equipment was mostly positioned "side-on" to windows, where this was not the case this was largely dictated by room layout (***Figure 7***).

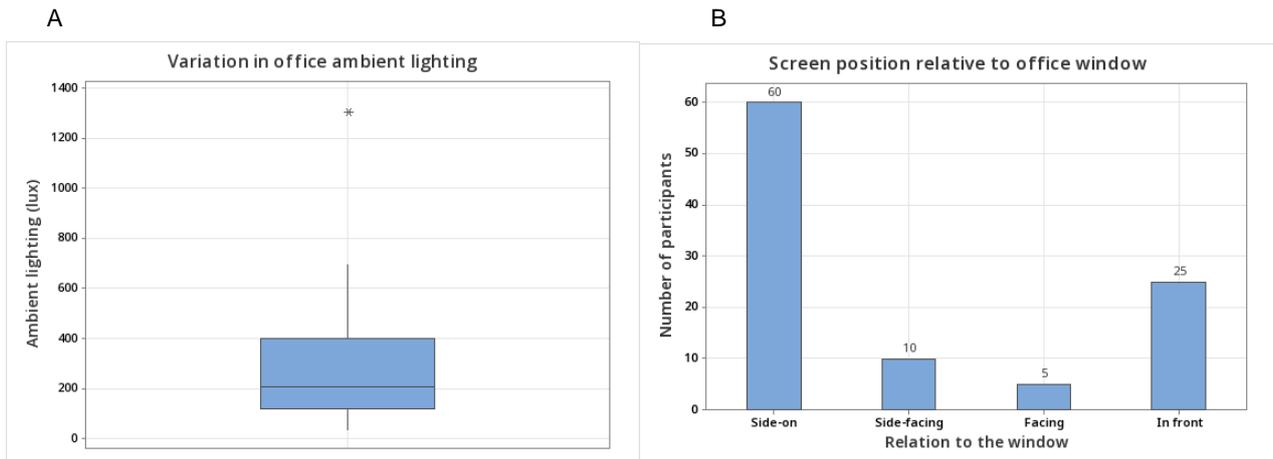

*Figure 7. Figure 7. The ambient lighting (A) and screen positioning (B) varied amongst participants as shown*

**Pathologists preferences**

There was notable variation in the pre-experiment settings on the displays, ranging from 120 – 345 cd/m2 (0-100 on the monitor backlight scale setting). As part of the deployment, these monitors were installed with a standardised default backlight setting. This indicates that pathologists, or those supporting pathologists with their display set up, have previously adjusted the backlight intensity.

Microscope light preferences ranged from 0.06 – 5.2 lux, representing nearly a 100-fold difference in preference. Pathologists generally had quite narrow preference ranges on the microscope (16/20 had IQ range <0.5 lux), but a few exhibited a much broader range (P6; IQ range 2.2 lux) (***Figure 8***).

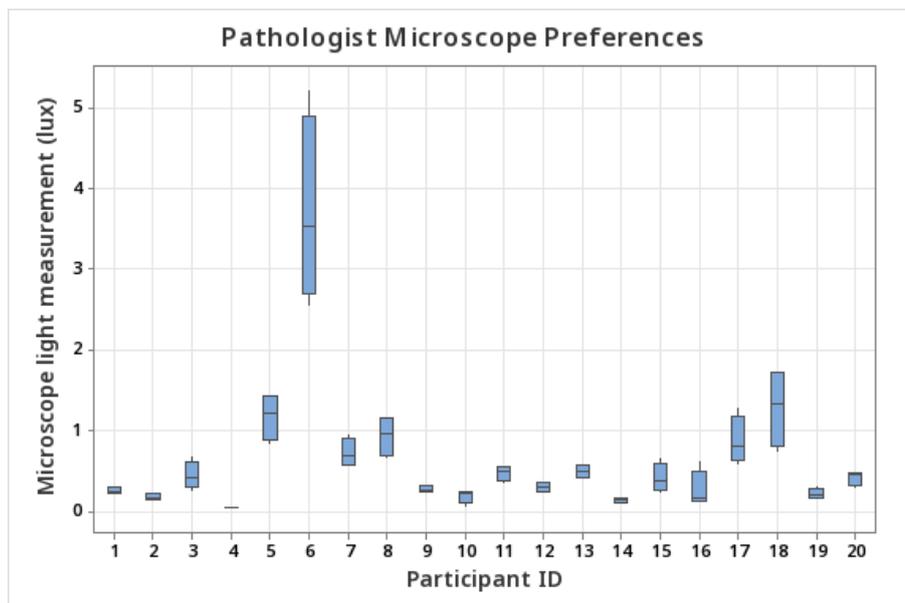

*Figure 8. Pathologists' microscope light preferences for viewing the "test" slide of breast tumour tissue.*
Display luminance preferences on the digital display ranged from 120 - 345cd/m$^2$, representing a more limited inter-pathologist variation than seen at the microscope. However, a similar pattern was observed whereby some participants had relatively narrow interobserver variability for preference (P8; IQ range 21.25 cd/m$^2$) and others showed less distinct preference (P18; IQ range 245.50 cd/m$^2$) (*Figure 9*). This finding was supported by comments of a few participants during the screen part of the experiment who felt that they would be happy at a range of light levels on their displays and found it harder to define their preference point.

Using average preference values for display luminance, 70% of pathologists prefer a screen luminance of 250cd/m$^2$ or less, 85% 300 cd/m$^2$ or less and 100% 500cd/m$^2$ or less.

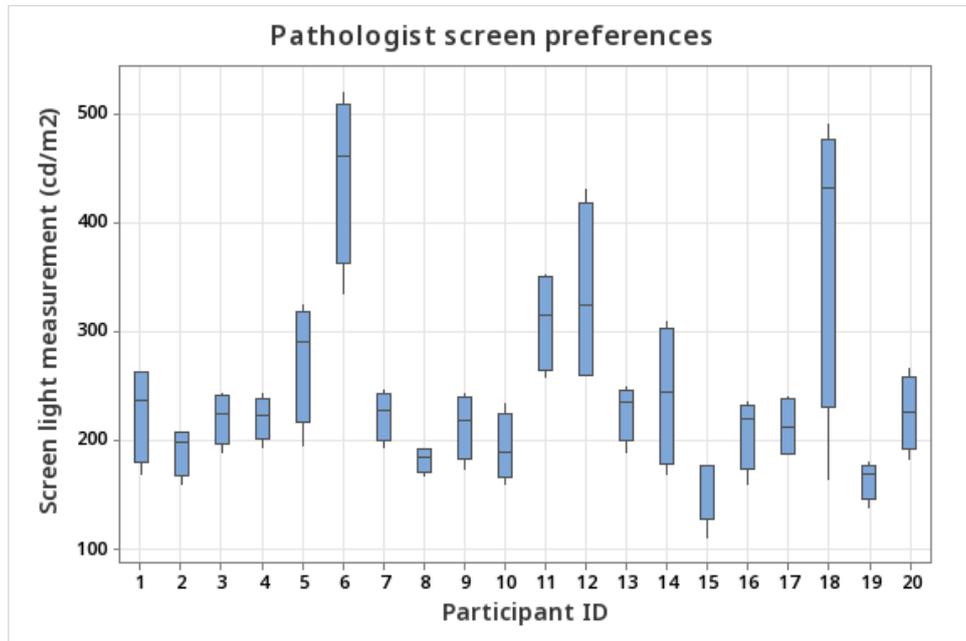

*Figure 9.* Pathologists' display screen luminance preferences for viewing the "test" image of breast tumour tissue.

**Preference correlation**

In general, no correlation was seen between microscope and display preference. One participant had a markedly higher microscope light preference, and this did correlate with a high screen luminance preference. If excluding the two 'high brightness preference' participants (P6 and P18) r=0.163, including them skews the data and r=0.75 The screen preferences also showed no correlation with ambient lighting (Correlation coefficient of 0.27).

For the majority of the participants, the screen preferences expressed represented relatively minimal changes from their "pre-experiment" settings. However, for some users a marked increase or reduction was made (*Figure 10*). This is supported in our survey finding that the majority of users are happy with their current display luminance.

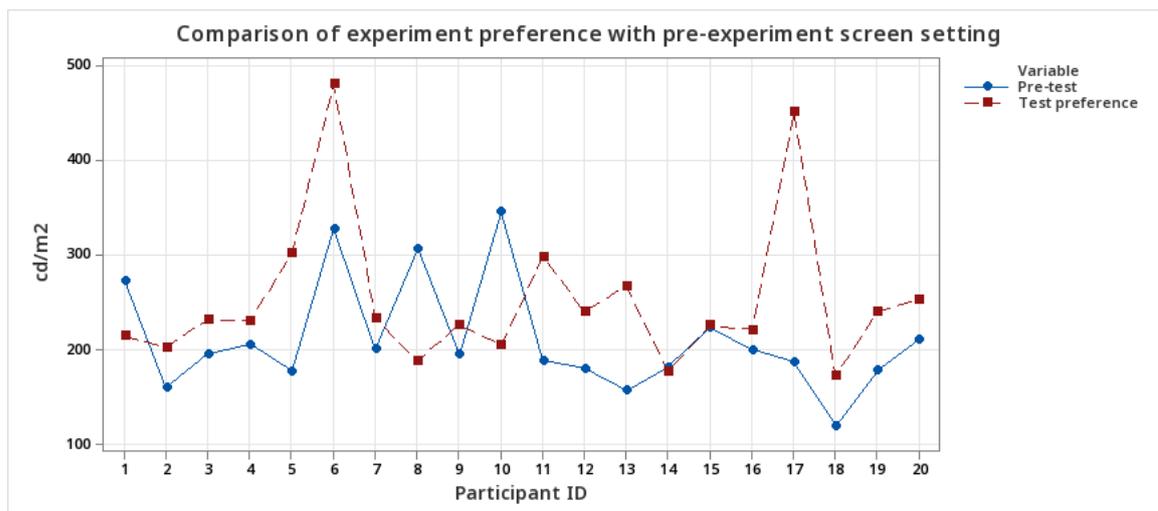

*Figure 10.* Pre-experiment measured screen luminance compared with measured preference (mean) during the experiment.

**Experiment comments**

A number of the participants made comments that they felt their preference would vary on different days for a number of reasons such as tiredness or room lighting. It was also noted by a few that their tolerated range at the display was probably quite broad and that a diagnostically acceptable range may be broader still.

A few of our participants have extensive digital pathology experience and amongst these colleagues, it was commented that different display luminance preferences have emerged – akin to the variation seen across microscope users (P15). Increasing the luminance of the display was described as helpful for the diagnostic image but after a longer period of reporting becomes too uncomfortable and tiring (P8). Another colleague also commented that mitoses are easier to identify when the screen backlight is increased (P5).

**DISCUSSION:**

Our results confirm the anecdotally observed difference in light use between pathologists at the microscope. Analyses of multiple aspects of pathologist reporting activities have been conducted previously [18–20] but this is the first attempt to describe light use as far as we are aware. The accompanying survey provides richer context by capturing the wide variation in frequency of adjustments and the multifactorial reasons for doing so as well as providing opportunity to gather pathologist opinions on the importance of light adjustments for their work.

Results from the light preference task indicate that pathologist preferences are present. For many these are consistent within a relatively narrow range. Whilst preference ranges are generally wider for the screen, our results indicate that a "one size fits all" approach to display set up will not suit all pathologists.

The lack of correlation between microscope and screen preference means microscope preference cannot be used to guide screen set up, except perhaps where users have a very bright microscope preference as this was better correlated with a brighter screen preference. Lack of correlation may relate to differences in what users are trying to achieve with light adjustment on each modality, as highlighted by our survey results in which microscope adjustments were predominantly for slide factors whereas for the screen adjustments for environmental lighting and visual comfort were more common.

For departments looking at procurement of digital displays – it useful to note that preferences of all pathologists in this study were less than 500cd/m$^2$ and that 85% preferred screen luminance of 300 cd/m$^2$ or less. This was surprising given previous work in this department where a range of screens were evaluated by pathologists for digital pathology viewing and the highest luminance monitor (up 2100cd/m$^2$) was preferred [5].

The light preference task was conducted in consultants' offices, where there was variation in ambient light levels, but surprisingly screen preferences were not correlated to ambient light. Several studies in radiology have demonstrated ambient light has an impact on performance [21–23] so further assessment is needed to understand the relationship between preference and performance in pathology. Such studies led to control of ambient light in reporting environments of radiologists to achieve sufficient contrast detection without inducing visual strain from screen use at high luminosity [11]. The practicality of low ambient light working for pathologists seems limited until the whole workflow is digital given the number of other tasks pathologists undertake when reporting [18] but should be considered as an option to achieve visual comfort for those with lower light preferences or who are experiencing visual strain. Current pathology workspaces should also be altered to optimise screen position relative to bright light sources such as windows, as this has been shown to reduce performance in a pathology specific contrast performance task, with the optimal position being side-on to a window [24]. However, alterations may not be possible in all cases. In this study, some pathologists were limited in adjusting their equipment layout by room constraints and to ambient light levels by the absence of blinds.

The overall strength of opinion on the importance of light adjustment on both modalities indicates that this is a function that may need to be addressed in digital display development. Whilst the number who adjust their screen light settings was small, a range of frequency of adjustments was represented in this sample. Current monitors are far less easy than a microscope to adjust and this was captured in our survey, with some pathologists not sure how to do so. Adoption of new technologies is linked with perceived ease of use [25] and in this situation it may be achieved by offering an experience that is familiar and as easy as microscope light adjustments. Display development in other areas has sought to use this principle and shown increased speed and comfort with digital pathology when image viewing software captures the feel of microscope slide review [26,27]. These results may therefore be of interest to display developers and manufacturers. Facilitating preference needs to be balanced against performance and technical image parameters may need to be defined within which users can make fast, "eyes free" adjustments as they work.

Some of the pathologists in our survey were less sure about how light use would impact reporting digitally, highlighting a lack of knowledge of displays in general as well as inexperience in digital pathology – a knowledge gap also recently described by Abel et al [6]. The need to adjust for ambient lighting and visual comfort is a sufficient argument for this function in the "new microscope" but it was useful to capture the thoughts of more experienced digital pathology users who similarly to at the microscope find that short term light adjustments can help identify specific features such as mitoses more easily. The development of clear guidance and education on screens and their functions will be needed to provide pathologists with confidence in the use and adjustment of their reporting equipment, especially if pathologists are to be responsible for it as recent FDA guidance suggests [6]. The development of tools such as the Point of Use Quality Assurance tool (a free online tool in which users check their performance in contrast detection test) can support pathologists to assess their own working environment [28].

Strengths & Limitations

Our survey was conducted by a large sample (64 pathologists across 6 NHS trusts) with a good response rate (59%) and respondents across the full age range, who represent a mix of experience (trainees and consultants) and working hours. Responses in relation to microscope light use habits and opinions should generalise well. The variation in digital pathology experience of this cohort means screen light use habits and opinions may be less generalisable.

The light preference task took place in large pathology department with established digital pathology workflows and good exposure to digital pathology, however use of digital pathology for primary reporting in this cohort was variable and was not captured in our results, except where users made comments specifically relating to their experience. Less experienced users may have found identifying their preference on this modality more challenging and thus it may be a less accurate reflection of the variation of preference.

A real-world approach was used in the light preference task, accepting lack of control for several variables such as environment lighting, equipment positioning, use of light filters on microscopes, other screen or image viewer software adjustments and testing at different times of the day or week. We felt this approach most suited our research aim of being better able to set up users for digital pathology in our department.  Further work which addresses the impact of these parameters may help to define optimal working conditions for digital pathologists.  Similarly, it would be valuable to understand how light use and preference changes during a prolonged period of reporting rather than the snapshot we have captured here. As supported by our survey, visual comfort is a common precipitant of light adjustment across both modalities and this becomes more relevant over time, where accommodation and visual fatigue are more apparent [29]. Finally, whilst we strongly feel there is a need to define critical performance parameters of these digital pathology screens, this work shows that individuals, their preferences and their working habits need to be considered in the process.

**COMPETING INTERESTS & ACKNOWLEDGEMENTS**

David Brettle has provided consultancy to Jusha Commercial & Trading Co, Ltd on medical displays.

Thank you to all the participating pathologists across the West Yorkshire Association of Acute Trusts Region who generously contributed their time in completing the survey and/or practical assessment.

**ETHICS**

Requirement for ethical oversight in this study was waived on review by Leeds Teaching Hospitals NHS Trust Research and Innovation department. Subjects in this study gave informed consent for their participation. Samples of human tissue were used in generating the slides and images used in part of this work. Informed consent was not obtained for this, as these were surplus tissue samples remaining after diagnosis and used as control tissue, they were fully anonymised, and their use without specific consent was approved by a research ethics committee (Leeds West LREC reference 05/01205/270)

**FUNDING**

National Pathology Imaging Co-operative, NPIC (Project no. 104687) is supported by a £50m investment from the Data to Early Diagnosis and Precision Medicine strand of the government's Industrial Strategy Challenge Fund, managed and delivered by UK Research and Innovation (UKRI).

**AUTHOR CONTRIBUTIONS**

CJ, DT and DB conceived and planned the study. CJ carried out the experiments and data acquisition. CJ and DB analysed the data and CJ wrote the manuscript, which was revised by DT and DB. The final version has been approved by all authors.

**DATA AVAILABILITY**

Raw results from both the online survey and practical light preference assessment can be made available on request from corresponding author CJ.

**Keywords:** digital pathology, microscope brightness, display luminance, ambient lighting, guideline, recommendations

**SUPPLEMENTARY MATERIALS**

S1 – Survey pathologist light use and preferences

Pathologists accessed the survey online via Microsoft forms using the following final questionnaire template.

WP3.2: Pathologist light preference_V6_CJ                    18/5/21

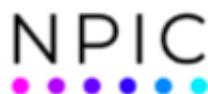  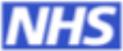

# Pathologist Light Preference
# Participant Questionnaire

This questionnaire is part of an NPIC project to study differences in pathologists' light preferences and usage, either when looking at slides digitally or on the microscope. The aim is to understand the requirements for working on a digital display for pathologists.

Your participation in this questionnaire is voluntary. You may choose not to participate and if you decide to participate, you may withdraw at any time. The questionnaire does not involve collecting uniquely identifying information and your responses will be kept confidential and stored safely.

The questionnaire will take around 5 minutes to complete.

**Please answer the following questions:**

1) How old are you?

25-34 ☐   35-44 ☐   45-54 ☐   55-64 ☐   65-74 ☐

2) What is your role?

Trainee ☐   Consultant ☐

3) If you are a trainee, what stage of training are you?

Stage A ☐   Stage B ☐   Stage C ☐   Stage D ☐

4) Approximately how many hours a week do you spend viewing slides on a microscope?

..............................................................................................................................

5) When looking at slides on a microscope, do you ever adjust the microscope light?

Yes ☐   No ☐   *(If no, please move to question 8)*

6) How often do you estimate that you adjust your microscope light settings?

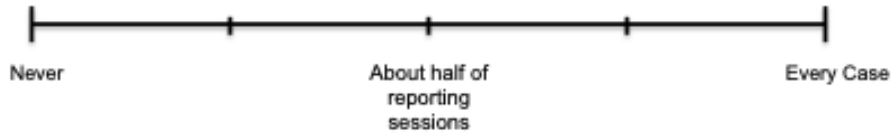

Never         About half of reporting sessions         Every Case

7) If yes to Q5, what reasons may precipitate this adjustment? (*Please tick all that apply and comment to explain why if possible*)

To examine specific features in a slide ☐         For particular tissue types ☐
Viewing slides at different objectives ☐           For specific stains ☐
To adjust for slide quality issues ☐               Visual comfort ☐
Changes in the environment lighting ☐              To adjust from another user ☐
Other (please specify below) ☐

..............................................................................................................
..............................................................................................................
..............................................................................................................
..............................................................................................................
..............................................................................................................

8a) Approximately how many hours a week do you spend looking at pathology images digitally?

..............................................................................................................

b) What is your predominant use for digital pathology? *Please circle*

   None    MDT    Primary reporting    Secondary opinions    Research    Teaching/Training

9) How do you find the brightness of your monitor?

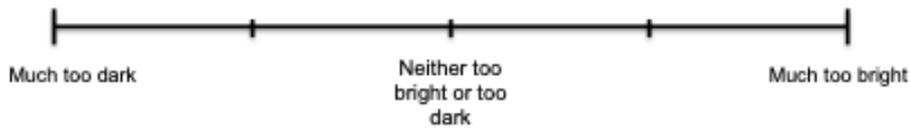

Much too dark         Neither too bright or too dark         Much too bright

10) When looking at digital images on a screen, do you ever adjust the screen backlight?

Yes ☐   No ☐   (*If No, please* move to question 12)

10) How often do you estimate that you adjust your screen brightness/backlight settings?

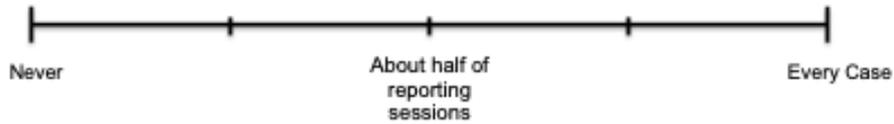

Never | About half of reporting sessions | Every Case

11) **If yes to Q9, what reasons may precipitate this adjustment?**
*Please tick all that apply and comment to explain why if possible*

- ☐ To examine specific features in a slide
- ☐ Viewing slides at different objectives
- ☐ To adjust for slide quality issues
- ☐ Changes in the environment lighting
- ☐ To adjust from another function (eg word document)
- ☐ For particular tissue types
- ☐ For specific stains
- ☐ Visual comfort
- ☐ To adjust from another user
- ☐ Other (please specify below)

......................................................................................................................
......................................................................................................................
......................................................................................................................
......................................................................................................................
......................................................................................................................
......................................................................................................................

12 **Please indicate to what extent you agree or disagree with the following statements?**
*Please tick one option per statement*

|  | Strongly Disagree | Disagree | Undecided | Agree | Strongly Agree |
|---|---|---|---|---|---|
| It is important to be able to adjust the light on a microscope. |  |  |  |  |  |
| It is important to be able to adjust the light on a digital display. |  |  |  |  |  |

b) **Why?** *Please comment to explain the importance ratings you gave.*

......................................................................................................
......................................................................................................
......................................................................................................
......................................................................................................

THANK YOU

S2 – Development of microscope adapter for lightmeter

Light microscopes are illuminated using a complex sequence of lenses and diaphragms in a process called Köhler illumination which achieves uniform illumination of the sample and provides high sample contrast. This is achieved by ensuring the image of the light source is perfectly defocused in the sample plane and conjugate image planes ie. the illuminating rays pass parallel through the sample. The amount of light entering the sample can be controlled by the condenser diaphragm or by reducing power to the light source.

Additional eyepieces (commonly used for training to allow side by side reporting) or cameras (used to photograph a specimen) can be inserted before the main eyepiece and involve the use of additional lenses to direct the image forming rays and the illuminating rays. In our department there are many different microscopes – variable in their make, model and age. While most consultants have an attachment to allow double-headed viewing, trainees usually do not. A minority of the consultants also have a camera attachment. This means that the light intensity of the bulb in the base of the diaphragm (which could be measured from the collector lens is variably representative of the light intensity at the eye piece. For this reason, we wanted to devise a method to measure the light output from the eyepiece directly. In a review of the literature, we could not find any examples of others having measured this before.

An integrating light sphere (ILS) is a spherical cavity coated in a highly reflective material which allows incoming light to undergo multiple reflections so that the intensity of the light becomes uniform. Some part of this reflected light can then be measured by a detector placed at a port within the sphere. Integrating light spheres are used in a variety of settings and can measure many different light sources, however from a review of available products there was no existing model which would be practical or affordable for this project.

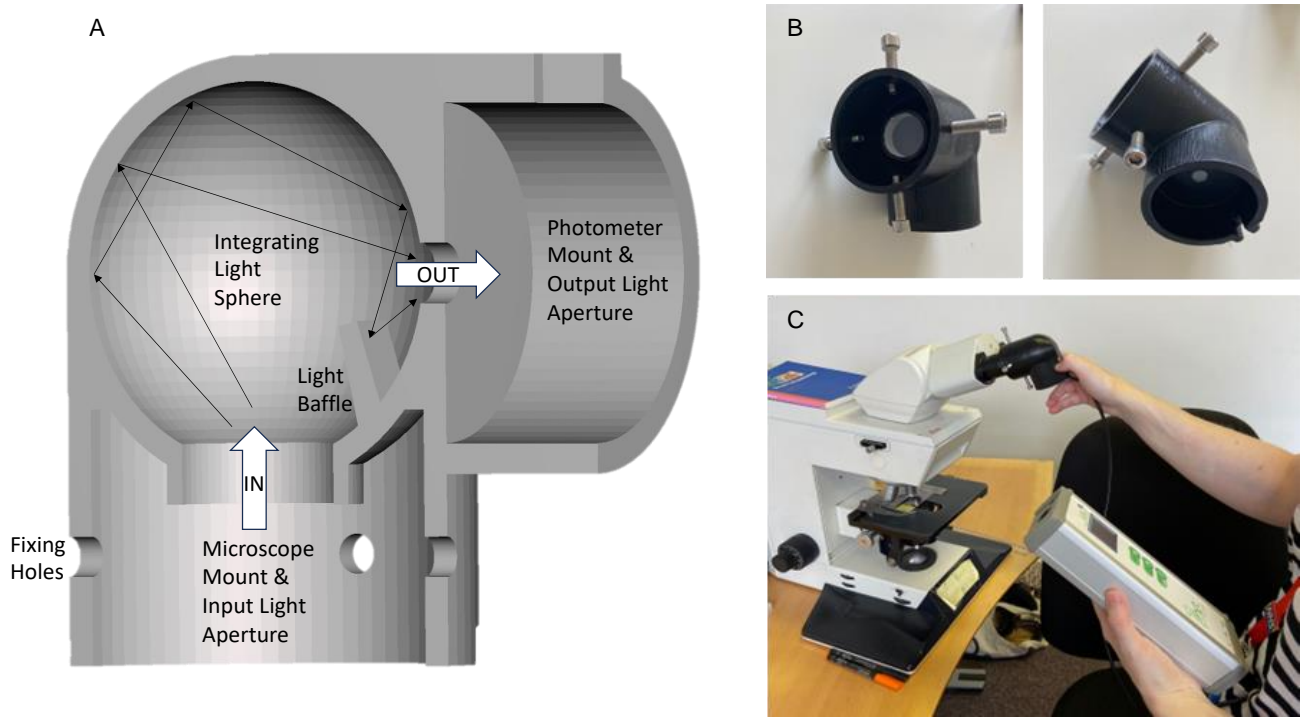

*Figure S1*: (A) CAD design of the integrating light sphere; (B) Images of the final integrating light sphere adapter; (C) Integrating light sphere in use, connecting to the microscope eyepiece and LXCan Spot luminance meter to measure the light output directly from the microscope.

Our searches identified work by Tomes and Finlayson (2016) to create a low cost integrating light sphere with a 3D printer for measurement of photoluminescence quantum yield [30] which was subsequently validated by da Cruz Junior & Bachmann (2021) [31]. We decided to use a similar approach to develop an integrating light sphere-like adaptation for our LXCan light meter. Key features of the design were portability (to allow use in multiple consultant offices), adaptability (to allow use with the range of microscopes in our department) and compatability with our existing light meter. Using the principles of an ILS our model was designed and tested.

Our focus was creating a tool which was able to reliably discriminate between fine increments of light adjustment at the microscope for intra and interpathologist comparisons and cross modality correlations, rather than achieving true accuracy of light measurement. The final iteration of this development process is pictured in supplementary *Figure S1* alongside images of design and the tool in use.